\documentclass[aps,prd,twocolumn,groupedaddress,showpacs,amsmath,amssymb]{revtex4-1}

\usepackage{graphicx}
\usepackage{bm,color}

\newcommand{\be}{\begin{eqnarray}}
\newcommand{\ee}{\end{eqnarray}}
\newcommand{\nn}{\nonumber\\}
\newcommand{\la}{\langle}
\newcommand{\ra}{\rangle}

\begin{document}


\title{Quantum simulation of $(1+1)$-dimensional U(1) gauge-Higgs model on a lattice \\
by cold Bose gases}


\author{Yoshihito Kuno$^1$}
\author{Shinya Sakane$^2$}
\author{Kenichi Kasamatsu$^2$}
\author{Ikuo Ichinose$^1$}
\author{Tetsuo Matsui$^2$}
\affiliation{$^1$Department of Applied Physics, Nagoya Institute of Technology, Nagoya, 466-8555, Japan}
\affiliation{$^2$Department of Physics, Kindai University, Higashi-Osaka, 577-8502, Japan}


\date{\today}

\begin{abstract}
We present a theoretical study of quantum simulations of 
$(1+1)$-dimensional U(1) lattice gauge-Higgs models, which contain 
a compact U(1) gauge field and a Higgs matter field, 
by using ultra-cold bosonic gases on a one-dimensional optical lattice. 
Starting from the extended Bose-Hubbard model with on-site and nearest-neighbor 
interactions, we derive the U(1) lattice gauge-Higgs model as a low-energy 
effective theory. The derived gauge-Higgs model exhibits nontrivial phase 
transitions between confinement and Higgs phases, and we discuss the 
relation with the phase transition in the extended Bose-Hubbard model. 
Finally, we study real-time dynamics of an electric flux 
by the Gross-Pitaevskii equations and the truncated Wigner approximation.
The dynamics is governed by a bosonic analog of the Schwinger mechanism, 
i.e., shielding of an electric flux by a condensation of Higgs fields, 
which occurs differently in the Higgs and the confinement phase.
These results, together with the obtained phase diagrams, shall guide experimentalists
in designing quantum simulations of the gauge-Higgs models by cold gases. 
\end{abstract}

\pacs{11.15.Ha,	
67.85.Hj	
}


\maketitle

\section{Introduction}{\label{intro}}
To understand physics of a given quantum many-body system in a coherent manner 
is not an easy matter. For example, the time development of these systems have not 
been clarified enough because high dimensionality of the state-vector space brings 
great difficulty for solving many-body Schr\"odinger equation \cite{Noack}. 
The sign problem also prevents us from carrying out 
straightforward Monte Carlo simulations for models including fermionic matter fields \cite{Troyer}. 

In the last several years, as an alternative approach, 
quantum simulations of various quantum systems by using ultra-cold atomic gases 
on an optical lattice have attracted lots of interests \cite{coldatoms}.
This comes mainly from high controllability and versatility of such atomic systems.
Quantum simulations of systems under a synthetic gauge field have already
become feasible due to experimental developments \cite{mg_optical}.
Furthermore, recent studies have proposed several ideas for realizing simulators of 
quantum systems involving {\it dynamical gauge fields} \cite{WieseZohar}.
Gauge systems appear in various fields of physics as an effective model, and
to understand their dynamics is one of the most important problems.
Also, proposals for an experimentally feasible atomic simulator of gauge systems 
still remain as an open problem.

In the previous papers \cite{U1GHM,NJP1,Future}, some of the present authors 
discussed how to simulate lattice gauge-Higgs models (GHMs) \cite{Fradkin,ichimaturev} 
by bosonic atoms on a $D$-dimensional optical lattice 
($D=3$ in Refs.~\cite{U1GHM, Future} and $D=2$ in Ref.~\cite{NJP1}).
The starting model in that approach is a cold atomic system described by 
the extended Bose-Hubbard model (EBHM) \cite{Baier,Dutta}; a lattice model of bosons having    
both on-site and off-site interactions. The target gauge model, GHM,
is a U(1) lattice gauge theory \cite{wilson} involving a compact U(1) gauge field
and a Higgs field in the London limit.
Reflecting the nonrelativistic nature of the EBHM,
interactions in the GHMs are asymmetric in the space and time directions;
some couplings in the spatial directions have no partners in the time direction
in contrast to the lattice gauge theory considered in high-energy physics \cite{wilson}.
However, the GHMs have a very interesting phase diagram as we shall see, and quantum simulation of the GHMs is expected to give us a very useful insight on the dynamics of 
the gauge theory.

In this approach \cite{U1GHM,NJP1,Future}, 
the phase degree of freedom $\hat{\theta}_a$ 
of boson operator $\hat{\psi}_a=\exp(i\hat{\theta}_a)\sqrt{\hat{\rho}_a}$ on the 
site $a$ of the optical lattice plays the role of the compact U(1) gauge field 
$\hat{\theta}_{r,i}$ defined on the link $(r,r+\hat{i})$ of a $D$-dimensional (spatial) lattice
of lattice gauge theory ($i =1,\cdots,D$ is the direction index). 
Compared with the alternative approach called quantum link model
(gauge magnet) \cite{quantumlink}, this approach has an advantage to express
a compact U(1) gauge theory in terms of  
the genuine U(1) operator  $\exp(i\hat{\theta}_{r,i})$ \cite{advantage}.
The Hamiltonian of the EBHM is known to contain
an annoying term that breaks Gauss law of the pure gauge theory 
(i.e., theory containing no matter fields) and
forces one to make a severe fine-tuning of its coefficient to generate a local gauge 
symmetry \cite{WieseZohar}.
However, inclusion of the Higgs field converts this breaking term to a 
term expressing the genuine Gauss-law with Higgs charge and relieves us from 
the fine-tuning.  

In this paper, we consider theoretically the atomic quantum simulation of the 
GHM with $D=1$ , supposing a system of Bosonic atoms 
on a one-dimensional (1D) optical lattice. 
This work is motivated by the following reasons: 

(i) From experimental point of views, quantum simulation 
of gauge theories in a 1D spatial lattice is expected to be 
easier than that in higher-dimensional lattices \cite{WieseZohar}. 
The present approach is based on a set of single-component atoms with long-range interaction (e.g., dipole-dipole interaction) loaded on a 1D optical lattice.
It has the following advantages \cite{advantage};  
first, this set up is proved to be feasible experimentally \cite{1Doptical_DDI}.
Second, a digital quantum simulation of the Schwinger model has been realized 
recently by using a 1D chain of several sites of trapped ions \cite{Martinez}.
In contrast, the present 1D optical lattice is scalable, i.e., 
hundreds of sites with large number of atoms can be prepared.
Third, some approaches such as Ref.~\cite{Bazavov} use the quantum link model,
 whose realization requires two-component atoms on the optical lattice.
The single-component atoms in the present approach is certainly more feasible 
in experiments. 

(ii) We can expect nontrivial behavior of the phase transitions of the target GHM, 
not found in the higher-dimensional cases \cite{U1GHM,NJP1,Future}. 
As shown later, the phase transition is governed by the Berezinskii-Kosterlitz-Thouless 
(BKT) type transition \cite{BKT}. 

(iii) Non-equilibrium physics of the lattice gauge theories has not been deeply understood yet. 
This subject is closely related to the recent heavy-ion collision experiments in 
high-energy physics \cite{HIC}. In our previous paper \cite{NJP1,Future}, we applied 
the Gross-Pitaevskii (GP) model \cite{GP} to study the real time dynamics of an electric flux, 
showing that the dynamics qualitatively realizes the behavior expected in each 
phase of the GHM. On the other hand, since the quantum fluctuations become 
more important in 1D system, we should include some quantum correction 
to the GP approach. Actually, real-time dynamics of a lattice gauge model has been studied by 
tensor network method \cite{TNN}, which is mainly applicable for 1D quantum systems with 
low filling. Here, we use the truncated Wigner approximation \cite{TWA} to 
study the effects of quantum fluctuations. 

Let us explain the basic steps of the present paper. 
First, by applying the method in our previous papers \cite{U1GHM,NJP1,Future} 
to the 1D EBHM {\it for large fillings of atoms} 
(large average occupation numbers of atoms per site) we obtain the (1+1)D GHM 
with short-range interactions. Here the additional +1D in the GHMs represents the 
imaginary-time axis in path-integral quantization for which we also use a lattice regularization. 
Next, we calculate the phase diagram of the derived GHM, which contains the Higgs phase 
and the confinement phase separated by the BKT transition. 
This phase diagram is compared with that of the EBHM to discuss the interesting relations 
between the strongly-correlated bosons and the gauge systems; namely, the Higgs phase 
corresponds to the superfluid phase and the confinement phase to Mott-insulator phase. 
This also provides an estimation of the relevant parameter region to simulate the 
phase transitions in experiments. 
Finally, we study the real-time dynamics of the GHM by using the GP 
equation \cite{GP} and the truncated Wigner approximation \cite{TWA}. 
This mean-field approach naturally arises because of the assumption of the large fillings 
of atoms. Numerical results show that a prepared electric flux exhibits quite different
behaviors depending on the Higgs or confinement phase. A phenomenon similar to the 
Schwinger mechanism is observed numerically. 

The paper is organized as follows. 
In Sec.~\ref{relation}, we present a derivation from 
the experimentally realized model, 
the EBHM on the 1D optical lattice,
to  the target simulated model, the (1+1)D U(1) GHMs
which follows our previous works \cite{U1GHM,Future}. 
In Sec.~\ref{phasediagram} 
we discuss the phase diagram of the GHM calculated by the Monte Carlo simulations. 
The obtained phase diagram is compared with that of the EBHM, whose details of the 
calculation is described in Appendix~\ref{appa}.
In Sec.~\ref{dyn}, we study non-equilibrium physics of the GHM, especially focusing on 
real-time dynamics of an electric flux. We present conclusion in Sec.~\ref{concle}

\section{Relation between the extended Bose-Hubbard model and the gauge-Higgs model}\label{relation} 

\begin{figure}[t]
\centering
\includegraphics[width=7cm]{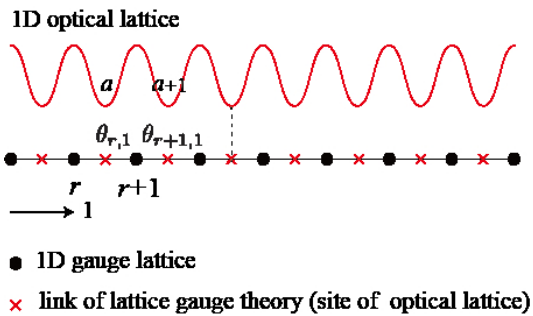}
\caption{(Color online) Optical and gauge lattices.
Each site $a$ of the optical lattice corresponds to a link of the gauge lattice
as $a\leftrightarrow (r,r+1)$. On the gauge link $(r,r+1)$ sits the spatial
component $\theta_{r,1}$ of gauge field.
}
\label{fig.lattice}
\end{figure}

We start with the EBHM defined on a 1D optical lattice (see Fig.~\ref{fig.lattice}). 
The hamiltonian $H_{\rm EBH}$ is given as
\begin{equation}
H_{\rm EBH}=\sum_a\left[-J(\hat{\psi}^{\dagger}_a\hat{\psi}_{a+1}
+\hat{\psi}^{\dagger}_{a+1}\hat{\psi}_a)  
+{U\over 2}\hat{\rho}^{2}_a+V\hat{\rho}_a\hat{\rho}_{a+1}\right],
\label{BHM}
\end{equation}
where $\hat{\psi}^{\dagger}_a$ and $\hat{\psi}_a$ 
are creation and annihilation operators of 
bosonic atoms on the site $a$, respectively, satisfying 
$[\hat{\psi}_a,\hat{\psi}^\dag_{a'}]=\delta_{aa'}$, etc. 
By representing $\hat{\psi}_{\alpha} = e^{i \theta_{\alpha}} \sqrt{\hat{\rho}_{\alpha}}$, 
we have the phase operator $\hat{\theta}_{\alpha}$ 
and the density operator $\hat{\rho}_{\alpha}$.
The coefficient $J$ represents the hopping strength, $U$ the on-site interaction, 
and $V(>0)$ the nearest-neighbor (NN) replusive interaction generated by, e.g., 
a dipole-dipole interaction \cite{DDI}. 
The above on-site repulsion $U ( > 0)$ represents the sum of $s$-wave scattering interaction
$U_{s}$ and on-site dipole-dipole interaction $U_{d}$; $U= U_{s}+U_{d}$. 
The $s$-wave scattering amplitude $U_{s}$ is highly controllable by Feshbach resonance \cite{Feshbach} 
and the hopping $J$ is controlled by the strength of the laser that
makes the optical lattice. 
On the other hand, the NN repulsion $V$ is determined by the kind of the atoms
loaded on the lattice and also depends on the lattice spacing.
In particular, ratios $J/U$ and $V/U$ can become highly-controllable parameters 
in real experiments by changing lattice depth \cite{coldatoms} and $s$-wave scattering
lengths \cite{Feshbach}, respectively.

The procedure \cite{U1GHM,NJP1,Future} to derive the 
GHM with local interactions from the EBHM for homogeneous 
density average at large fillings is based on a couple of
assumption; (i) the average density is homogeneous and large,
$\rho_{0a} (= \la \hat{\rho}_a\ra) =\rho_0 \gg 1$ (large filling)
and (ii) the density fluctuation  
around $\rho_0$ is small. The point (ii) implies that we focus on the sector 
of low-energy excitations.  Explicitly, we write the density operator as
\begin{equation}
\hat{\rho}_a=\rho_0+\hat{\eta}_a,
\end{equation}
and expand $H_{\rm EBH}$ with respect to 
the fluctuation operator $\hat{\eta}_a$ up to the second-order.
This kind of expansion is widely used in atomic physics \cite{coldatoms} .

To regard this expanded Hamiltonian as that of the GHM, we introduce
a gauge lattice and a set of operators in gauge theory defined on it.
The gauge lattice is the dual lattice of the optical lattice  (see Fig.~\ref{fig.lattice});
the site $r$ of the gauge lattice sits on the middle point of the link $(a,a+1)$ 
of the optical lattice. We define the 1st component of the vector potential $\hat{\theta}_{r,1}$  
and its conjugate momentum, namely the electric field $\hat{E}_r$, 
on the gauge link $(r,r+1)$ as
\begin{equation}
\hat{\theta}_{r,1}=(-)^r\hat{\theta}_a,\quad
\hat{E}_r=-(-)^r\hat{\eta}_a,
\label{A_E}
\end{equation}
where we use the same character $\theta$ both for atomic and gauge variables
(They can be distinguished by whether it carries the direction suffix 1).
Strictly speaking, the electric field should carry the direction suffix 1 as 
$\hat{E}_{r,1}$, but we omit it in this paper because there are no other components
in a 1D system. On the other hand, we keep the suffix 1 in $\hat{\theta}_{r,1}$
because we shall introduce the zeroth component of vector potential later 
[see $\theta_{x,0}$ in Eq.~(\ref{ZGH})].      
These pairs satisfy the canonical commutation relations \cite{wilson},
\be
[\hat{E}_r,\hat{\theta}_{r',1}] = -i\delta_{rr'},\ 
[\hat{\eta}_a,\hat{\theta}_{a'}] = i\delta_{aa'},
\ee 
which come from the commutation relations of $\hat{\psi}_a$.
The factor $(-)^r$ in Eq.~(\ref{A_E}) may look curious but is necessary to generate the Gauss law 
as we shall see.
Then the expanded Hamiltonian 
up to $O(\eta^2)=O(E^2)$ is expressed as
\begin{align}
H_{\rm EBH} &= C_{\rho_0} +H_{\rm GH}+ \Delta H+O(E^3),\nn
C_{\rho_{0}}&=\left( {U\over 2}+V \right)\sum_{r}\rho^{2}_0,\nn
H_{\rm GH}&= \sum_r\left[{V \over 2}(E_{r+1}-E_{r})^{2}+
\frac{g^2}{2}E^{2}_r\right. \nn
&\quad\ \  \left.-2J\rho_0\sum_{r}\cos(\theta_{r+1,1}+\theta_{r,1})\right], \nn
g^2&=U-2V,\nn
\Delta H&=\frac{J}{4\rho_0}\sum_{r}(E_{r+1}+E_{r})^2\cos(\theta_{r+1,1}+\theta_{r,1}),
\label{HGH}
\end{align}
where we  omitted hat symbols for operators.
There appear no terms linear in $E_{r}$ because $\rho_0$ is chosen
such that it gives rise to a stationary point of the energy.
$C_{\rho_{0}}$ is an irrelevant constant.
The hopping $J$ term of Eq.~(\ref{BHM}) generates both
the last term of $H_{\rm GH}$ and the term $\Delta H$.
Their coefficients are $J\rho_0$ and $J/\rho_0$ respectively, and the latter is   
suppressed by an extra factor $\rho_0^{-2}$ for large fillings.
Therefore we neglect $\Delta H$ and keep
$H_{\rm GH}$ as the Hamiltonian of the GHM for large fillings.

The first term of $H_\text{GH}$ leads to Gaussian with respect to $\nabla_1 E_{r} 
\equiv E_{r+1}-E_{r}$,
the divergence of the electric field, and gives rise to 
$\la(\nabla_1 E_r)^2 \ra\simeq V^{-1}$. 
This becomes the Gauss-law constraint $\nabla_1 E_r=0$ for 
{\em pure gauge theory} 
in the limit $V\to \infty$, and this limit is just the fine tuning required 
for the pure gauge theory in Refs.~\cite{WieseZohar}.
After introduction of a Higgs field in the London limit, this term represents
the genuine Gauss law ($\nabla_1E_r=Q_{\rm Higgs}$)
with the Higgs charge $Q_{\rm Higgs}$
 (see Ref.~\cite{Future} for details).

\begin{figure}[b]
\centering
\includegraphics[width=7cm]{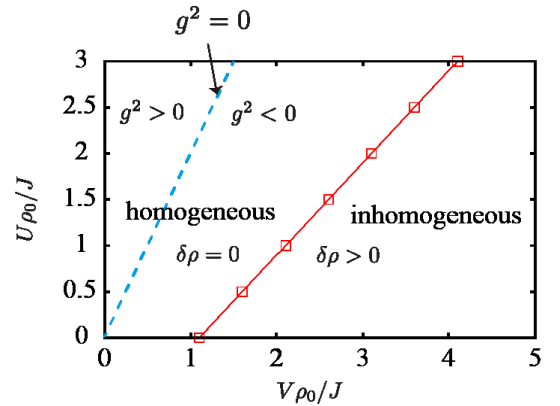}
\caption{
(Color online) Mean-field result for the 1D ground-state configuration 
of the average density $\rho_0$. 
It shows that the homogeneous configuration 
($\delta\rho\equiv \rho_{\rm even}-\rho_{\rm odd}=0$) is stable for $g^2>0$ as expected.
However, a homogeneous state exists even for $g^2<0$ for moderate
values of negative $g^2$. 
The inhomogeneous configuration stands for the density-wave pattern ($\delta\rho>0$). 
Detailed study on this point is given in Appendix A. 
}
\label{Figdw1d}
\end{figure}

The second term represents the energy of electric field.
In the ordinary lattice gauge theory used in high-energy physics 
\cite{wilson}, its coefficient $g^2 $ is the square of so called real gauge coupling 
constant $g$  and $g^2$ is treated as positive definite.
In our case, $g^2=U-2V$ may take \textit{negative value} (although 
we used a suggestive notation $g^2$ that it may be positive).
Here we can see that a homogeneous configuration of $\rho_{0a}=\rho_0$ is realized 
for $g^2 > 0$ by using a very simple type of the mean-field theory, i.e., 
two site mean-field theory. 
We take only two nearest-neighbor site of the Hamiltonian $E_{\rm{EBH}}$, called even-site and odd-site. 
By putting a mean field ansatz $\psi_{\rm even} \rightarrow \sqrt{\rho_{\rm even}}$ and $\psi_{\rm odd}\rightarrow \sqrt{\rho_{\rm odd}}$
 into the two sites, the EBHM mean field energy $E_{\rm{EBH}}$, which depends on the two mean density is given as
\be
E_{\rm{EBH}}=-2J\sqrt{\rho_{\rm even}\rho_{\rm odd}}+\frac{U}{2}(\rho^{2}_{\rm even}+\rho^{2}_{\rm odd})+V\rho_{\rm even}\rho_{\rm odd}\nonumber\\.
\ee
Then by detecting the energy minimum under a canonical ensemble constraint, $\langle\rho_{a}\rangle=\rho_{0}$ $(\rho_{\rm even}+\rho_{\rm odd}=2\rho_{0})$, 
we can show the homogeneity of the atom density. 
Fig.~\ref{Figdw1d} indicates the homogeneity of the atom density in ($U$-$V$)-plane.
Here, we found that,
$\rho_{0a} $ is homogeneous for $g^2 > 0$, i.e., $\rho_{0a}=\rho_0$, while
for $g^2 \ll 0$, $\rho_{0a} $ takes an inhomogeneous
DW configuration such as
$\rho_{0a}=\rho_0+(-)^a\delta\rho/2$ (see Fig.~\ref{Figdw1d}).
Result in Fig.~\ref{Figdw1d} also indicates the existence of a homogeneous
`phase' even for $g^2<0$.
This comes from the fact that the `critical line' $g^2=U-2V=0$ is 
renormalized by the hopping $J$-term in $H_{\rm EBH}$ in Eq.~(\ref{BHM}).
In Appendix A, we give rather detailed discussion on this homogeneous state
and obtain the global phase diagram of the EBHM. 

The third $J\rho_0$ term of $H_{\rm GH}$ expresses the NN coupling of the U(1) gauge field
$U_{r,1}=\exp(i\theta_{r,1})$ as $J\rho_0 U_{r+1,1}U_{r,1}$+ H.c., 
but it breaks the gauge symmetry under
$\theta_{r,1}\to \theta_{r,1}+\lambda_{r+1}-\lambda_r$
($\lambda_r$ is an arbitrary $r$-dependent real parameter) \cite{wilson}.
In Refs.~\cite{U1GHM,NJP1,Future} we introduced
a U(1) Higgs field $\phi_r=\exp(i\varphi_r)$ 
in the London limit (its amplitude is frozen to unity) on the site $r$,
and regard this term 
as a gauge-fixed version of the Higgs coupling to the unitary
gauge $\phi_r=1$ as 
\be
&&-J\rho_0\sum_r\phi^\dag_{r+2}U_{r+1,1}U_{r,1}\phi_r +{\rm H.c.}\nn
&\rightarrow& 
-J\rho_0\sum_rU_{r+1,1}U_{r,1}
+{\rm H.c.}
\ee
This Higgs coupling and the remaining terms in $H_{\rm GH}$
is invariant under the local U(1) gauge transformation \cite{wilson}
\be
&&U_{r,1}\to U'_{r,1}=V_{r+1}U_{r,1}\bar{V}_r,\ \ \phi_r\to \phi'_r=V_r\phi_r,\nn 
&&E_{r}\to E'_{r}=E_{r}, \ \ V_r\equiv\exp(i\lambda_r),
\label{gtrsf}
\ee
where
the bar symbol in $\bar{V}_r$ denotes the complex conjugate.
In short, $\tilde{H}_{\rm GH}(\phi, U, E)\equiv H_{\rm GH}(\phi U\phi^\dag, E)$
is a gauge-invariant Hamiltonian which we formally work with 
and the gauge invariant observables are calculable by its gauge fixed version $H_{\rm GH}(U,E)$
of Eq.~(\ref{HGH}). 

Here we comment that, in the present 1D system, the Gauss law term of gauge theory
is to be supplied  only by the $U$- and  $V$-terms in $H_{\rm EBH}$.
This is in contrast with the higher dimensional [(2+1)D and (3+1)D] systems 
in which one must take additional next-NN interactions into account \cite{NJP1,Future}.
This point shows that the 1D atomic system is more feasible 
to realize a quantum simulator in experiments than the higher dimensional systems.

To derive the gauge model,  we start with the partition function
$Z_{\rm GH}={\rm Tr}\exp(-\beta\tilde{H}_{\rm GH}(\phi,U,E))$  defined
from $H_{\rm GH}$ of Eq.~(\ref{HGH}) 
with the inverse temperature $\beta = 1/(k_{\rm B}T)$ 
expressed in path-integral in canonical formalism
(sum over canonical pair of coordinate and momentum).
Then all the fields  are  defined on the (1+1)D lattice with its site denoted by $x=
(x_0,x_1)\equiv(\tau,r)$
as $\phi_x, U_{x,1}, E_{x}$,
where $\tau=0,\cdots, N-1$ is the site  index along the imaginary-time axis
discretized by the spacing $\Delta \tau \equiv \beta /N$ and
we take the limit $N \to \infty$ formally.
In this paper we omit the hat symbol in $\hat{\mu}\ (\mu=0,1)$ to denote the unit
vector in the $\mu$-th direction. Therefore $x+0$ and $x+1$ imply the NN site
$(\tau+1,r)$ and $(\tau,r+1)$ respectively. We use  $\nabla_\mu$ 
as the forward difference operator $\nabla_\mu f_x \equiv f_{x+\mu} -f_x$.

We introduce the imaginary-time ($\tau$) component of the gauge field  
$\theta_{x,0}$ on the link $(x,x+0)$ 
to make the first Gauss-law term of $H_{\rm GH}$ a linear form in $\nabla_1E_{x}$,
i.e.,
\begin{eqnarray}
&&\exp\Big[-\Delta\tau{V\over 2}(E_{x+1}-E_{x})^2\Big] \nonumber  \\
&&\hspace{0.5cm}
=\int d\theta_{x,0}\exp\Big[-{1 \over 2V\Delta\tau}
\theta^2_{x,0}+i\theta_{x,0}\nabla_1E_{x}\Big].
\end{eqnarray}
Because every step in the approach
of Refs.~\cite{U1GHM,NJP1,Future} is applied in the present (1+1)D system 
in a straightforward manner, we present the expressions for 
$Z_{\rm GH}$ as (For details see Sec.~II of Ref.~\cite{Future}),
\be
\hspace{-0.5cm}
Z_{\rm GH}&=&\int[d\theta_1][dE]\nn
&&\times\exp\left[\sum_\tau\left(
iE_{x}(\theta_{x+0,1}-\theta_{x,1})-\Delta\tau H_{\rm GH}\right)\right]\nn
\hspace{-0.5cm}
&=&\int[d\theta_0][d\theta_1][d\varphi]\exp(A_{\rm GH}),\nn
\hspace{-0.5cm}
[d\theta_\mu]\!&=\!&\!\prod_x\!\frac{d\theta_{x,\mu}}{2\pi},\
[d\varphi]\!=\!\prod_x\!\frac{d\varphi_x}{2\pi},\ [dE]\!=\!\prod_x\!\sum_{E_{x}=-\infty}^{\infty}.
\label{ZGH} 
\ee
The action $A_{\rm GH}$ may be expressed in terms of the compact U(1) variables 
$U_{x,\mu}=\exp(i\theta_{x,\mu})$ on the link $(x,x+\mu), \mu=0,1$  $[x=(x_0, x_1)=(\tau,r)]$ and 
$\phi_x=\exp(i\varphi_x)$ as
\be
A_{\rm GH}&=&A_{\rm I}+A_{\rm P}+A_{\rm H},\nn
A_{\rm I}&=&\frac{c_1}{2}\sum_{x}\bar{\phi}_{x+0,}U_{x,0}\phi_{x}
+{\rm c.c.}, \nn
A_{\rm P}&=&\frac{c_2}{2}\sum_{x}\bar{U}_{x,0}\bar{U}_{x+0,1}U_{x+1,0}U_{x,1}
+{\rm c.c.},\nn
A_{\rm H}&=&\frac{c_3}{2}\sum_{x}\bar{\phi}_{x+2,}U_{x+1,1}U_{x,1}\phi_x
+{\rm c.c.}.
\label{AGH} 
\ee
The parameters in $A_{\rm GH}$ are expressed in terms of those in $H_{\rm GH}$
and $\Delta \tau$ as
\begin{equation}
c_1={1 \over V\Delta\tau}, \; \ 
c_2={1 \over g^2\Delta\tau}, \; \
c_3=2J\Delta \tau\rho_0.
\label{cs}
\end{equation}

The action $A_{\rm GH}$ in Eq.~(\ref{AGH}) contains only the short-range interaction,
and each term is obviously invariant under a 
{\em time-dependent local gauge transformation}
[given by Eq.~(\ref{gtrsf}) with the replacement $r\to x =(\tau,r)$],
and the term $A_I$ is 
nothing but the gauge-invariant kinetic term of the Higgs field 
$\phi_x$ \cite{Fradkin}.
The present action (\ref{AGH}) is derived from the nonrelativistic
Hamiltonian (\ref{BHM}) and has no relativistic invariance under change of
directions $\mu=0\leftrightarrow \mu=1$, which is in contrast with 
the conventional models \cite{wilson} in lattice gauge theory.
However, this model supports both the confinement and Higgs phases (see Sec.~\ref{phasediagram}) 
and therefore is interesting enough for simulation of the gauge theory.

\section{Phase diagram of the GHM}\label{phasediagram}

In this section we calculate the phase diagram of the GHM defined by
Eqs.(\ref{ZGH}) and (\ref{AGH}) by MC simulations.
For ordinary second-order phase transitions, 
one may use the specific heat $C_{\rm GH}$,
\be
C_{\rm GH}&=&\frac{1}{L^2}\la (A_{\rm GH}-\la A_{\rm GH}\ra)^2\ra,
\label{sh}
\ee
to locate the transition point by the peak of $C_{\rm GH}$.
However, because the present GHM is a (1+1)D system with a global U(1) symmetry as we shall see, 
the general theorem allows neither second-order transition nor long-range order.
One may expect, if any, a phase transition of the same type as the BKT transition 
that is well known in the 2D XY spin model \cite{BKT}.  
This expectation is plausible because the original 1D EBHM itself is known to have a phase 
transition at zero temperature between the Mott-insulator and superfluid, 
where the superfluid has  a quasi-long range order due to the low dimensionality of the space.

Let us summarize the 2D XY spin model and its BKT transition. Its action is given by
\be
A_{\rm XY}=J\sum_x \sum_{\mu=0,1}\vec{S}_{x+\mu}\vec{S}_{x}=
J\sum_{x,\mu}\cos(\sigma_{x+\mu}-\sigma_x),\nn
\ee
where $\sigma_x \in [0,2\pi)$ is the XY spin angle defined on $x$ and
$\vec{S}_x\equiv(\cos\sigma_x,\sin\sigma_x)$.
The BKT transition of the 2D XY model  is of infinite order, signaled 
by divergence of the susceptibility $\chi$,
\be
\chi\equiv\sum_y\la \vec{S}_y\vec{S}_x\ra =\sum_y\la\cos(\sigma_y-\sigma_x)\ra,
\label{sus}
\ee
at the transition point $J=J_c\simeq 1.12$. 
This value deviates from the round peak of specific heat 
by $\sim 20 \%$. 
The spin-spin correlation function
$G(x)\equiv \la \vec{S}_x\vec{S}_0\ra$ for large $|x|$ behaves exponentially
$\exp(-|x|/\xi)$ in the spin disordered ($J < J_c$) phase and  
algebraically $|x|^{-\eta}$  in the spin quasi-ordered ($J > J_c$) phase.
The correlation length $\xi$ and the critical exponent $\eta$ depend on $J$ as
$\xi(J) \propto \exp(b/\sqrt{J_c-J})$ near $J_c$  and $\eta(J_c)=1/4$.

\begin{figure*}[t]
\centering
\includegraphics[width=13cm]{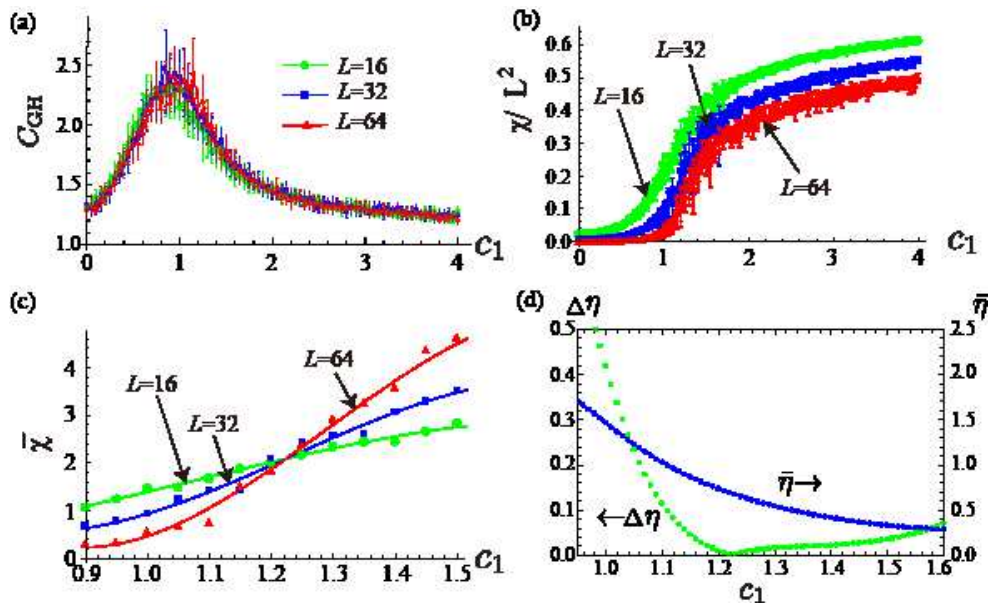}
\caption{(Color online) Thermodynamic quantities of the GHM of Eq.~(\ref{ZGH})
and their scaling behavior for $c_2=c_3=2.0$: (a) Specific heat $C_{\rm GH}$ of Eq. (\ref{sh}).
(b) Susceptibility density $\chi/L^2$ of Eq.~(\ref{sus}).
(c) Scaled susceptibility density $\bar{\chi}$ of Eq.~(\ref{ssus})
with the choice of critical values $c_1= c_{1c} \sim 1.22$ and $\eta(c_{1c}) \sim 0.68$.
These values are determined so that the three curves of $\bar{\chi}$ for $L=16,32,64$ 
merge at a single point (see text). 
A large value of $\eta(c_{1c})$ reflects the fluctuations of  $\theta_{x,0}$ in $A_{\rm I}$, which
suppress the correlations among $\{\sigma_x\}$. 
(d) Average critical index $\eta(c_1)$ and deviation $\Delta\eta(c_1)$ from a
triple intersection (see the text for their definitions).
$\Delta\eta$ decreases as $c_1$ increases and reaches  its minimum $\Delta\eta = 0$
at $c_{1c}$ (three curves merge there) and keeps small values. 
Thus $c_{1c}$ is interpreted as a BKT-type critical point dividing the disordered phase ($c_1 < c_{1c}$)
and quasi-ordered phase ($c_1 > c_{1c}$) as Eq.~(\ref{ssus}) shows.  }
\label{fig_GHM}
\end{figure*}

To locate the phase transition point, one may observe the scaling behavior 
of $\chi$ of Eq. (\ref{sus}) for lattice of size $L\times L$ with 
the periodic boundary condition.
{\it In the quasi-ordered phase}, power decay of $G(x)$ gives rise to 
$\chi \propto L^{2-\eta(J)}$  for large $L$. 
One may define the scaled susceptibility density as $\bar{\chi}(J,\eta^*)\equiv
L^{\eta^*-2}\chi(J)$ with arbitrary parameter $\eta^*$. 
Then one has
\be
\hspace{-0.5cm}
\bar{\chi}(J,\eta^*)\!\equiv\! L^{\eta^*-2}\chi(J)\!\propto\! L^{\eta^*-2}\!\times\! L^{2-\eta(J)}
\!=\!L^{\eta^*-\eta(J)}.
\label{ssus}
\ee
Therefore, at $\eta^*=\eta(J)$, $\bar{\chi}(J,\eta(J))\propto L^0$ becomes scale invariant (no $L$-dependence).  
We note that this scale invariance holds  for every point of $J (> J_c)$.
Among possible multiple set of scale-invariant points $(J,\eta(J))$, the smallest value of $J$ is 
just the critical point $J_c$. 
On the other hand, {\it in the disordered phase},  behavior of
$\chi(J)$ depends on the relation between $L$ and $\xi$.  
For $L \gg \xi$ (i.e., $J$ is well below $J_c$), the exponential decay of $G(x)$ gives rise to 
$\chi \propto \xi^2$ and $\bar{\chi}(J,\eta^*)\propto  L^{\eta^*-2}\xi(J)^2$,
which shows no scale-invariant points are possible (for $\eta^*\neq 2$).
For other case ($L \lesssim \xi$), no explicit formula of $\chi(J)$ seems available.
Although there is no analytic proof  that $\bar{\chi}(J,\eta^*)$ has no scale invariant points 
throughout the disordered phase, we confirmed that it is the case 
by numerical analysis applying the same method to obtain Fig. \ref{fig_GHM}(d).

Below we shall see  a close relation between the GHM and the 2D XY model, and 
explore the possibility of a  BKT-type phase transition of the GHM.
Let us first 
consider the limit  of $c_1\to \infty$.  Then
$A_{\rm I}$ of Eq.~(\ref{AGH}) forces $U_{x,0}= 1\ [\theta_{x,0}=0$(mod $2\pi$)] 
up to gauge rotation.
By introducing the gauge-invariant angle variable $\sigma_x$ (we use the same letter as 
the above XY spin angle) on the spatial link 
$(x,x+1)$\ [i.e., between the site $(x_0,x_1)$ and $(x_0,x_1+1)$] as
\be 
\sigma_x &\equiv& (-)^{x_1} (\varphi_{x}+\theta_{x,1}-\varphi_{x+1}),
\label{sigma}
\ee
$A_{\rm GH}$ reduces up to a constant to
\begin{equation}
A_{\rm GH}\to
c_2\sum_{x}\cos(\sigma_{x+0}-\sigma_x)+
c_3\sum_{x}\cos(\sigma_{x+1}-\sigma_x).
\label{AGH2DXY} 
\end{equation}
This is just the action of the 2D XY spin model with the asymmetric couplings
$c_2$ and $c_3$. 
It has the global U(1) symmetry under $\sigma_x\to\sigma_x+\alpha$.
This limiting model exhibits BKT transitions on a certain critical line in the $c_2$-$c_3$ 
plane \cite{asymXY}.

As $c_1$ is reduced from infinity, $\theta_{x,0}$ is allowed to fluctuate more. 
However, it is plausible to expect that the above critical line in the $c_2$-$c_3$ 
plane survive for finite but large enough $c_1$, and the BKT-type scaling relation 
(\ref{ssus}) holds. Then, we determine the transition point $c_1=c_{1c}$ for a given set of
$(c_2,c_3)$ using the scaled susceptibility density $\bar{\chi}$ of Eq.~(\ref{ssus}) 
calculated by $A_{\rm GHM}$ and Eq.~(\ref{sigma}). 
Explicit procedures are as follows;\\
(i) Fix $c_2$ and $c_3$ and measure the susceptibility $\chi(c_1,L)$ along $c_1$ for 
three $L$'s ($L_1, L_2, L_3$).\\
(ii) Obtain interpolating curves    
$y(c_1,L)$ for $\chi(c_1,L)$ in third-order polynomials in $c_1$;\\
(iii) For $\bar{\chi}(c_1,\eta,L) =L^{\eta-2}y(c_1,L)$  (below we use $\eta$ instead of 
$\eta^*$ because no confusions may arise) , set the conditions
of a triple intersection;
$\bar{\chi}(c_1,\eta, L_1)=\bar{\chi}(c_1,\eta, L_2)=\bar{\chi}(c_1,\eta, L_3)$,\
and solve  these two equalities numerically. \\
From Eq. (\ref{ssus}), one expects
a continuous (multiple) set $(c_1,\eta)$ of solutions for every value of $c_1$
starting at and larger than the critical value $c_1=c_{1c}$.

In Fig.~\ref{fig_GHM}, we show (a) $C_{\rm GH}$ and  (b) susceptibility density 
$\chi/L^2$ for $c_2=c_3=2.0$ and $L_1=16$, $L_2=32$, $L_3=64$. 
Instead of multiple set of solutions, we find a unique triple intersection 
at $c_1 =c_{1c}\simeq 1.22$ for $\eta \simeq 0.68$.
In Fig.~\ref{fig_GHM} (c) we show three $\bar{\chi}$ with this value.
To understand the absence of multiple solutions,
we introduce the distance $\Delta\eta$ from the triple interaction defined as follows;
Each pair of three curves $\bar{\chi}(c_1, \eta, L)$ (e.g., $L_a$ and $L_b$) 
with fixed $c_1$ has an intersection  at $\eta=\eta_{ab}$.  
We define $\Delta\eta \equiv $ Max of three distances, $|\eta_{ab}-\eta_{a'b'}|$. 
In Fig.~\ref{fig_GHM} (d) we show $\Delta\eta$ together with the average of $\eta$,
$\bar{\eta} \equiv (\eta_{12}+\eta_{23}+\eta_{31})/3$.
The triple intersection at $c_1=c_{1c}$ locates at the unique minimum of $\Delta\eta\
[\Delta\eta(c_{1c})=0]$ and separates the large $\Delta\eta$-region ($c_1 < c_{1c})$ 
and the small $\Delta\eta$-region ($c_{1c} < c_1$). By regarding 
$\Delta \eta$ in $c_{1c} < c_1$ almost vanishing, Eq.~(\ref{ssus}) 
tells  us that  the point $c_1=c_{1c}$ is an approximate critical point 
of BKT-type phase transition separating the disordered phase ($c_1 < c_{1c}$)
and the quasi-ordered phase $(c_{1c} \leq c_1$).
Slow but steady development of $\Delta \eta $ from zero with increasing  $c_1 (> c_{1c})$ 
shows that there exist some corrections to the power-law decay of spin correlations.
One may account for it 
by finite-size effect generally; in particular, by 
the logarithmic corrections \cite {janke} that may become relatively important for smaller
$\eta$. 
Then, we expect that $\Delta \eta$ for $c_{1} > c_{1c}$ tends
to small if we use the date obtained for systems
with larger $L$.
To calculate the critical point  for general $(c_2, c_3)$ below, we repeat to identify 
a unique triple intersection  (scale-invariant point)  as the BKT-type critical point.
This manipulation is supported because a quite similar behavior of $\Delta\eta(J)$ is observed 
for the 2D XY model giving rise to $J_c\sim1.11(1.12), \eta(J_c)\sim0.23(0.25)$ in good agreement with the known results in parentheses for larger $L$.

\begin{figure}[b]
\centering
\includegraphics[width=7cm]{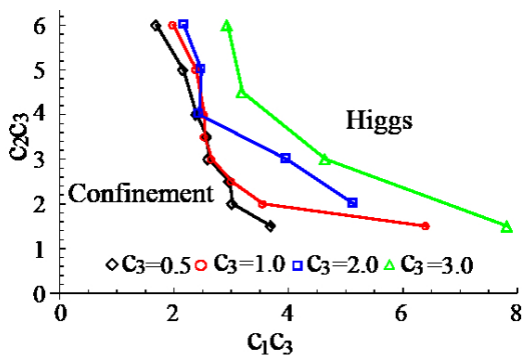}
\caption{(Color online) Phase diagram of the short-range GHM of Eq.~(\ref{ZGH})
(corresponding to EBHM for homogeneous large fillings) obtained by the path-integral
MC simulation.
In the Higgs phase, fluctuations of gauge field $\Delta\theta$ is small,
while,  in the confinement phase, $\Delta \theta$ is large
and fluctuations of the electric field $\Delta E$ is small.
}
\label{fig_PDGHM}
\end{figure}

In Fig.~\ref{fig_PDGHM}, we show the phase diagram in the ($c_1c_3$-$c_2c_3$) plane
obtained in the same procedure as in  Fig.~\ref{fig_GHM} (c).
The four transition lines correspond to $c_3=0.5,1.0,2.0$ and $3.0$.
The two coordinates $c_1c_3$ and $c_2c_3$ of this plane are chosen so that 
they are dimensionless and $\Delta \tau$ independent combinations 
(see Sec. III of Ref.~\cite{Future} for detailed discussion on this point). 
We note that there are no transitions in the region of $c_1 c_3\lesssim 1.4$ because
the data fitting in the form of Eq.~(\ref{ssus}) failed there. This is consistent with
our assumption that finite but sufficiently large $c_1$ is a necessary condition for
BKT-type transition.

Because the disordered phase necessarily realizes for small $c_1, c_2, c_3$'s, 
the lower (smaller $c_2c_3$) region  in Fig.~\ref{fig_PDGHM} corresponds to the disordered phase, and 
the higher (larger $c_2c_3$) region corresponds to the quasi-ordered phase.
In gauge theory, typical and well-known phases are Higgs, confinement, and Coulomb
phases. The magnitude $\Delta\theta$ of fluctuations of gauge field $\theta_{x,\mu}$ 
increases in this order \cite{wilson,Kogut,Fradkin}. 
The magnitude $\Delta E$ of fluctuations of conjugate electric field $E_x$ 
decreases in this order due to the uncertainty principle. Because we have two phases 
in  Fig.~\ref{fig_PDGHM}, we naturally identify the spin-disorder phase as 
the confinement phase and the quasi-ordered phase as the Higgs phase. 
This is supported by the phase diagrams 
obtained for related GHMs derived from the 2D and 3D 
EBHM \cite{U1GHM,NJP1,Future}.

\begin{figure}[t]
\centering
\hspace{-1cm}
\includegraphics[width=6cm]{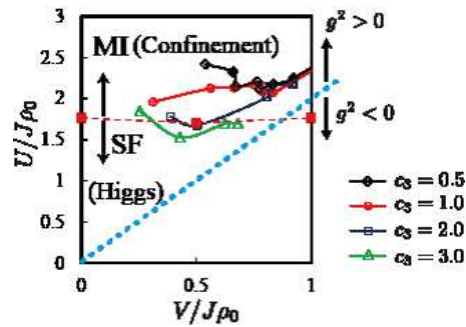}
\caption{(Color online) Phase diagram obtained by 
the path-integral MC simulations for \textit{L}=16, 32, and 64, 
where $L^{2}$ is size of the space-time lattice. 
There are two phases, SF (superfluid) and MI (Mott insulator). 
We show the gauge-theoretical interpretation of each phase. 
Experimental setup with the parameter regions labeled `Higgs' and
`confinement' are suitable for an atomic quantum simulation of the U(1) GHMs.
The blue broken line represents the vanishing gauge coupling, $g^2(\equiv U-2V)=0$.
The red dotted line is the phase boundary obtained in Appendix A.2.
From the view point of the gauge theory, 
for $g^2>0$ and large $U$ the ground state is confinement phase, 
in which the system has no phase coherence, whereas
for $g^2>0$ and small $U$ the ground state is Higgs phase, 
in which the system has strong phase coherence, i.e, SF phase. 
}
\label{fig.PD2}
\end{figure}

In order to identify the relevant parameter regions in experiments, 
it is important to compare the phase diagram of Fig.~\ref{fig_PDGHM} 
to that of the original EBHM Eq.~(\ref{BHM}). 
In Fig.~\ref{fig.PD2}, we show the phase diagram of the EBHM in small $V/J$ regime, which is obtained 
by the MC simulations in Appendix \ref{appa}.
There are two phases. In the Mott insulator (MI) phase, the strong on-site repulsion $U$ stabilizes
the uniform distribution of atoms without the phase coherence.
For the large hopping $J$, the superfluid (SF) forms as a result of 
the Bose-Einstein condensation. 
In Fig.~\ref{fig.PD2}, the red dotted line shows the phase boundary 
of the EBHM between the MI and the SF, 
which was determined by a scaling method.
For details, see Appendix A and Fig.~\ref{fig.PD3}.
To compare the phase diagrams of BHM and GHMs, we transport critical lines shown in
Fig.~\ref{fig_PDGHM} to those in the ($U/J\rho_{0}$-$V/J\rho_{0}$) plane of Fig.~\ref{fig.PD2}, by using the relations between the parameters such as
 $c_{1}c_{3}=2(V/(J\rho_{0}))^{-1}$ and 
 $c_2 c_3 = 2(J\rho_0)/(U-2V)$. The phase boundaries of the two models are fairly in good agreement for $g^2>0$ and 
this result certifies the existence the confinement-Higgs phase transition. 

In other words, our target phases are the MI phase and the SF phase in 
relatively small $V$ regime; the two phases in cold atomic system has explicit meaning of 
the lattice gauge theory. 
Fig.~\ref{fig.PD2} also shows suitable parameter regions for the quantum-simulation 
experiment on the GHM. 
This fact certainly supports that the confinement-Higgs phase transition can be observed
by experiments of ultra-cold atoms in a 1D optical lattice.

The phase diagram of the BHM in large $V$ regime itself has various interesting 
phases \cite{DMRG1,Kawaki}. 
As the NN repulsion is getting larger compared to the on-site repulsion, 
the assumption of uniform density configuration $\hat{\rho}_{a} \to \rho_{0}$ 
is broken, and the density wave phase appears in the phase diagram, 
in which an imbalance in the density at the even and odd sites exists. 
The result of the MC simulation for large $V$ is also given in Appendix \ref{appa}.

\section{Real-time dynamics of electric flux string}\label{dyn}

In the high-energy physics, study of non-equilibrium 
dynamics of the gauge theories in Minkowski space-time 
has been a challenging problem. 
Its detailed study is going to be even more important 
inspired by recent heavy-ion collision experiments \cite{HIC}.
In this section, we study real-time dynamics of an electric flux
in the Higgs and confinement phases observed by the path-integral MC simulations 
in Sec.~\ref{phasediagram}.
To this end, we employ the GP equation, which is a reliable method 
to simulate dynamical behavior of a condensed fluid in condensed matter physics especially for the superfluid behavior \cite{GP2}. 
In general the GP equation is suitable for the SF (Higgs) phase,
 however in this section we extend the limitation, i.e., apply the GP equation to the MI (confinement) phase 
and focus on investigating qualitative behavior of both phases.
A similar problem of real-time dynamics of an electric flux 
was addressed in an atomic quantum simulation of the Schwinger model, the (1+1)D
quantum electrodynamics (QED), in Refs.~\cite{Schw,TNN}, where the tensor network simulation was used.
Interestingly enough, the calculations in Refs.~\cite{Schw,TNN} report qualitatively 
different behavior of the electric flux depending on values of the gauge coupling and 
mass of the electron. 

From Sec.~\ref{phasediagram}, the present $(1+1)$D GHMs have two different phases, the Higgs 
and the confinement 
phases, therefore study of dynamical behavior of the electric flux gives important 
and interesting insight into the gauge dynamics in each phase.
A similar problem for the higher-dimensional cases has been already addressed 
in our previous work \cite{NJP1,Future}, where we have obtained qualitative difference of flux string behavior for each phases.
In the present $(1+1)$D case, the direct comparison 
between the GHM in Eq.~(\ref{HGH}) 
and the Schwinger model is possible for the string dynamics.

The GP equation can be derived from the coherent path-integral of the EBHM 
as a saddle point equation.
When the Lagrangian $L$ is described as $L=-{{\sum }_{a}}i\psi _{a}^{*}(d{{\psi }_{a}}/dt)-H_{\rm EBH}$, 
the Euler-Lagrangian equation $\delta L/\delta \psi^{*}_{a}=0$ becomes the GP equation.
From this procedure, the GP equation for the EBHM is given by
\begin{eqnarray}
i\frac{1}{J}\frac{\partial \psi_a}{\partial t}&=&-(\psi_{a-1}+\psi_{a+1})+\frac{U}{J}\rho_a\psi_a\nonumber\\
&&+\frac{V}{J}(\rho_{a-1}+\rho_{a+1})\psi_a+\frac{\mu '}{J}\psi_a,
\label{GP}
\end{eqnarray}
where we have added chemical potential term $\mu'\psi_a$ by which
 the average density $\langle|\psi_{0}|^{2}\rangle=(1/N)\sum_a|\psi_a|^{2}=\rho_{0}$ is 
controlled.
In what follows, we set the equilibrium density $\rho_{0}= 1$
by putting $\mu'=\rho_{0}(U+2V)-2J$.
Here it should be remarked that this is a kind of the normalization
and results for other densities can be  obtained by a simple rescaling as
$\psi_a \rightarrow \sqrt{\rho_0}\psi_a$ with $(U,V)\rightarrow {1\over \rho_0}(U,V)$.
A system of $200$ spatial lattice sites with the free boundary condition is used 
and the time step  $\Delta t$ for the numerical calculation is set as 
$\Delta t= 10^{-4}$.
The time scale $t=1$ is $\sim$ 0.32 [msec] by using the energy scale 
$U/h\sim$ 500[Hz] in typical experiments.
In Eq.~(\ref{GP}), the dissipation term is not included, and therefore the total 
energy is conserved during the time evolution.

As an example of a non-equilibrium state of the system, 
we consider the state with a single electric flux with a length $R$, 
i.e., we set the electric flux string in the initial state 
and solve the GP equation numerically. 
Hereafter we use the gauge lattice label, $a\rightarrow (r,1)$.
Explicitly, we set the initial configuration as 
$|\psi_{r,1}|^{2}=[1+(-1)^{r}(0.1)]\rho_{0}$ for $r_{1}-R\leq r\leq r_{1}+R-1$ 
($r_{1}$ is the center of the string). 
On the other sites, $|\psi_{r,1}|^{2}=\rho_{0}$.
In experiments the above initial flux is created by the density modulation \cite{Wurtz}.
In our numerical simulation, we set the length of the electric flux $R=60$. 
In the confinement phase, this length is expected to be long enough for observing the 
typical quark-confinement phenomena \cite{Kogut,Smit}.  

\begin{figure}[t]
\centering
\includegraphics[width=5cm]{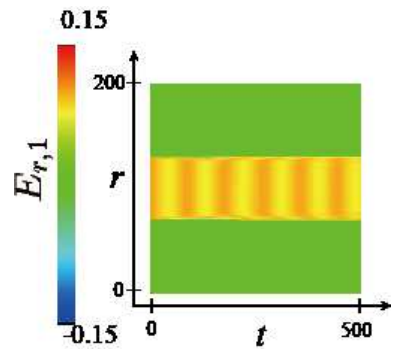}
\caption{Solution of the GP equation (\ref{GP}) for $U/J=10^{3}$, $V/J=1.0$
and ${g^2}/J=0.2$.
An electric flux string is stable because effect of the Higgs field is negligibly small.}
\label{GPsmallJ}
\end{figure}
\begin{figure*}[t]
\centering
\includegraphics[width=14cm]{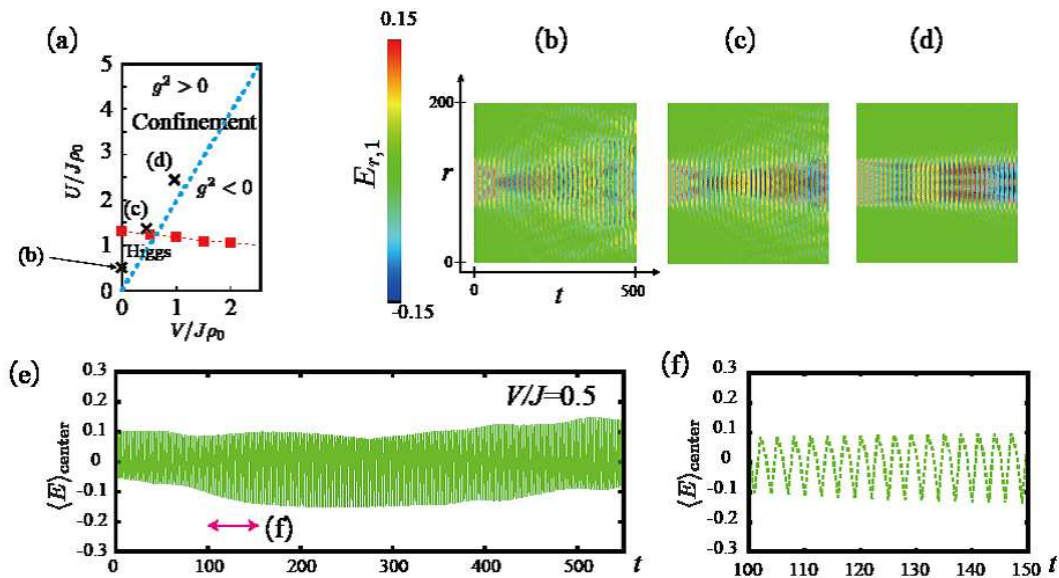}
\caption{(Color online) Real time-dynamics of a single electric flux.
(a): 
Global phase diagram of the EBHM  by the path-integral MC. 
(See Appendix A and Fig.~\ref{fig.PD3}). 
The black crosses in the phase diagram indicate the locations on which
the string dynamics is observed as in the right panels (b)-(d).
(b)-(d):
The time evolution of a single electric flux of length $R=60$ in $L=200$ lattice. 
All time evolutions have the same gauge coupling $g^{2}/J=0.2$.
For $V/J=0$ (b) and $V/J=0.5$ (c), the initial electric flux gradually spreads out 
along the time evolution and the pure-gauge Gauss law, div$E=0$, does not work. 
For $V/J=1$ (d), the shape of the initial electric flux is not broken. 
This indicates that the pure-gauge Gauss law works for the initial flux. 
For all cases (b)-(d), the electric fields oscillates changing their sign. 
This phenomenon is explained by the string-antistring oscillation 
that is also observed in the QED real-time simulation in Ref.~\cite{Schw,TNN}. 
(e) and (f): For the case (c), the electric field averaged
over the central ten sites, $\langle E \rangle_{\rm center}$. 
The oscillation period is $\sim 3.1$ [kHz] in our time scale and the period depends 
only on the value of $J$.}
\label{RD1}
\end{figure*}

We first briefly consider the case of sufficiently small $J$,
and verify that the electric flux string is stable by solving the GP equation
(\ref{GP}). In our numerical simulation, the electric field $E_r$ are defined as
\begin{eqnarray} 
E_r\equiv (-1)^{r}(|\psi_{r,1}|^{2}-\rho_{0}).
\end{eqnarray}
This corresponds to a density fluctuation from mean density.
The typical numerical results is shown in Fig.~\ref{GPsmallJ}.
When $J$ is very small, the effects of the Higgs field is very small and the system has
properties of the pure compact U(1) gauge theory.

In Fig.~\ref{RD1} (b)-(d), we show the time development of an electric flux 
along line ${g^2}/J=0.2$ 
in the phase diagram. 
The results show that from the Higgs phase to the confinement phase, the stability of 
the electric flux is increased.
In fact, there is a striking difference between the Higgs phase and the confinement phase; 
in the Higgs phase the electric flux spreads out. 
On the other hand in the confinement phase, the form of the initial flux does not change.
In addition, we find that in both phases an oscillation of the electric field in the flux takes place.
This behavior of the flux string means that a pair production of the Higgs particle 
and the resultant string breaking takes place. 
But in the confinement phase the restoration of the flux string readily occurs as a result of the strong 
gauge coupling. By contrast,
in the Higgs phase the initial electric flux is unstable, i.e.,
bits of electric flux appear and they spread out from the initial region
$r_{1}-R\leq r\leq r_{1}+R-1$.
This behavior comes from the condensation of the Higgs field and the resultant
spontaneous symmetry breaking of the global U(1) gauge symmetry, i.e., the charge
is not conserved in the Higgs phase. 

\begin{figure}[t]
\includegraphics[width=8cm]{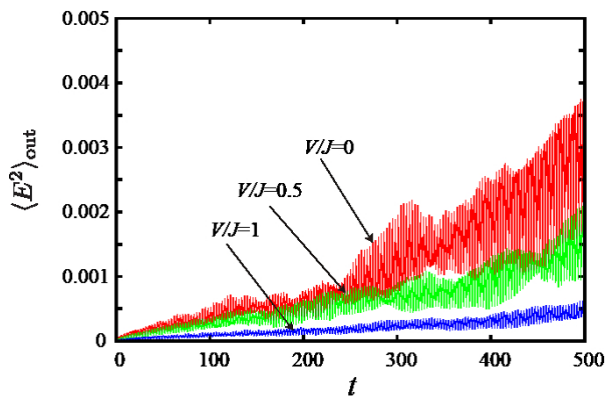}
\caption{(Color online) Mean value of  electric field per link 
out of the initial electric flux, $\langle E^2\rangle_{\rm out}$. 
As decreasing of the value of $V$, the electric field spreads out of
its initial string position.
${g^2}/J=0.2$.}
\label{div2}
\end{figure}

In order to evaluate the spread of the electric field from the initial string, 
we measure the mean value of  electric field strength 
$\langle E^2\rangle_{\rm out}$ outside of the initial place of the electric flux,
\begin{eqnarray}
\langle E^2\rangle_{\rm out}
\equiv\frac{1}{L-R}\sum_{ (r,1)\notin {\rm string \ line}}E^{2}_r.
\end{eqnarray}
Fig. \ref{div2} shows the $\langle E^2 \rangle_{\rm out}$ 
for some typical values of the Gauss-law coupling $V$ along the line ${g^2}/J=0.2$.
It is obvious that as the value of $V$ decreases,  $\langle E^2 \rangle_{\rm out}$ increases 
rapidly as a function of time, i.e.,
the initial configuration of the electric flux is unstable, and electric-field excitations
propagate out of the initial electric flux. 

\begin{figure}[t]
\includegraphics[width=7.5cm]{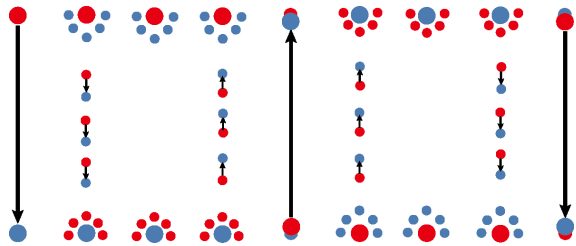}
\caption{(Color online) Schematic picture of string-antistring oscillation.
This phenomenon comes from the Schwinger mechanism that was first suggested 
for QED \cite{Schwinger}.
The red circle represents a positive charge particle and the blue one is negative charge particle.
The black arrow represents an electric flux tube between a pair of opposite
static charges.}
\label{schwinger_mec}
\end{figure}

In addition, Fig.~\ref{RD1} exhibits another remarkable behavior of the electric field
as non-equilibrium phenomena.
The results show that on the sites, where the initial electric flux exists, 
the electric fields are oscillating and changing their sign in both the Higgs and 
the confinement phases.
The data in Fig.~\ref{RD1} (e) and (f) shows the time evolution of the electric field averaged
over the central ten sites $\langle E\rangle_{\rm center}$. 
Similar behavior was observed in all three cases; the central electric field in 
all cases oscillates in time and takes even negative values.
The result is similar to behavior of an electric flux in Schwinger model with 
a small electron mass as reported in Refs.~\cite{Schw,TNN}, and 
the phenomenon is called the string-antistring oscillation.
Quantum link version of QED was also studied from this point of view \cite{Banerjee}.
All the above studies indicate that the pair creation takes place in the central 
region of 
the electric flux as the Schwinger mechanism tells \cite{Schwinger}.
The two initial static charges residing on the edges of the initial electric flux  
tend to be shielded by particles of opposite charge. 
The initial electric flux vanishes as a result. 
After that, the reverse process happens, i.e., the pair creation recurs and 
the electric flux in the opposite direction is created. 
Iterating this process causes consequently the string-antistring oscillation. 
Our results in Figs.\ref{RD1} (b)-(f) for the U(1) GHM 
suggest that a similar string-antistring oscillation process takes place as in QED.
Fig. \ref{schwinger_mec} illustrates schematically the string-antistring oscillation process.    

In the above, we studied the real-time dynamics of the electric flux string
by using the GP equation.
Results concerning to the Higgs phase are reliable as the Higgs phase is nothing 
but the SF and the order parameter $\psi_a$ has only small fluctuations there.
On the other hand, one may think that use of the GP equation for 
the MI regime is not appropriate even in the region close to the phase boundary 
to the SF. 
However, we expect that our study for the weak MI regime captures at least
qualitatively correct picture of the string dynamics.
In fact, the numerical results presented above are very similar to those obtained by
other numerical results for closely related gauge models in Refs.~\cite{Schw,TNN}.

Anyway, a more detailed study of the dynamics of the flux string in 
the confinement-MI 
regime is important and it gives some useful remarks on the experimental setup.
To this end, the idea of the truncated Wigner approximation (TWA)
 is useful \cite{TWA},
that is, we simulate the real-time dynamics of the flux string with different initial
conditions.
In the above, on solving the GP equation, we employed the initial condition 
such that the phase of the wave function $\theta_{r,1}=0$. 
However in the MI state, the phase of the `condensate' fluctuates as a result
of the uncertainty relation between the particle number and the phase.
In other words, the fluctuations of the phase cannot be avoided in the MI even if 
a prominent experimental technique to control the phase of the `condensate'
is used.
This fact arises questions about the reliability of the results obtained by the GP equation.

Here, to study the effect of the phase fluctuation, we employ the initial condition 
such that $\theta_{r,1}=\lambda_{r}\pi$ and solve the GP equation, where the variables
$\{\lambda_r\}$ are homogeneously distributed random numbers between 
$[-\Delta_{m},\Delta_{m}]$ with $\Delta_{m}>0$.
To calculate the physical quantities like the density at each site,
we take 100 samples, which have different initial conditions produced by the random
numbers $\{\lambda_r\}$, and take average of the calculated quantities over the samples. 
In this way, we obtain results of the real-time dynamics of the electric flux.
This sampling method essentially corresponds to the TWA \cite{TWA}. 

\begin{figure}[t]
\includegraphics[width=8cm]{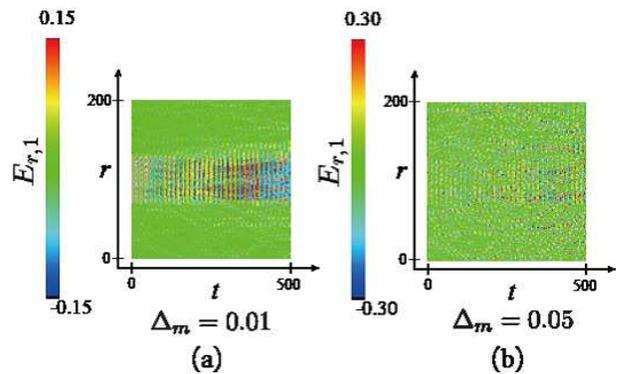}
\caption{(Color online) Real-time dynamics of the flux string in the confinement
region with random distributions of the wave-function phase in the initial condition.
These results were obtained by averaging 100 samples with different initial phase distribution.
For $\Delta_m=0.01$, the stable evolution of the string is observed, whereas
for $\Delta_m=0.05$, only small portion of the initial string configuration remains.}
\label{RD1_ran}
\end{figure}

Fig. \ref{RD1_ran} shows the real-time dynamics of the flux string for 
the parameters $V/J=1$, $U/J=2.2$, which locates in the weak MI region 
in the phase diagram, and for two cases of the random distribution of the phase,
$\Delta_{m}=0.01$ and $0.05$.
The other conditions are the same with those in the previous calculation.
For $\Delta_m=0.01$, the stable evolution of the string is observed, whereas
for $\Delta_m=0.05$, only small portion of the initial string configuration remains.
From the above result and the uncertainty relation of the number and phase, 
$\Delta n\cdot\Delta\theta \geq {1\over 2}$, the fluctuation of the atomic 
number $\Delta n$ at the initial state is required as 
$\Delta n\gtrsim 1/(0.05\cdot2\pi)\sim 3.2$.
To satisfy this requirement in experiments, the average particle number must
be sufficiently large and also a sufficient number modulation in the flux string
is needed.
If this condition is satisfied, the string-antistring oscillation in Fig.~\ref{RD1} 
is to be observed.
More detailed study on this problem by using the full TWA and its extension \cite{TWA}
will be reported in a future publication.

Finally let us comment on the real-time dynamics of the flux string in the 
confinement-MI region apart from the phase boundary to the SF.
In this region, the GP equation is not applicable, as the phase of the `condensate'
fluctuates rather strongly.
However, this region corresponds to the strong-coupling region of the GHM,
and therefore the string tension of the electric flux is strong.
Then it is naturally expected that the string is quite stable and its fluctuations
in time are small, and as a result its electric field is kept positive, $E_r>0$.
In fact, this phenomenon was observed in the numerical study of the gauge
models in Refs.~\cite{Schw,TNN}. 

\section{Conclusion}\label{concle}

We explained how to realize the atomic quantum simulation of the (1+1)D GHM by 
using cold atomic gases on the optical lattice. 
By means of the path-integral MC simulations, we obtained the 
phase diagrams of the GHM and also the EBHM at large filling and 
verified that these two phase diagrams are in good agreement with each other for 
the parameter region of the positive gauge coupling.
We examined the phases of the EBHM from the gauge-theoretical point of view 
and the result is summarized in Table \ref{EBHM_GHM}.
\begin{table}[tb]
\caption{ Correspondence between the phases of EBHM and GHM. 
$\Delta\theta$ is the magnitude of fluctuations of atomic phase $\theta_a$ 
(vector potential $\theta_{r,1}$).}
  \begin{tabular}{|c|c|c|c|} \hline
    EBHM & U(1) GHM & $\Delta \theta$ & $\rho_{0a}\equiv \la \hat{\psi}_a\ra$  \\ \hline
    MI &  Confinement & large & $\rho_{0}$ (homogeneous)\\ 
    SF & Higgs & small &  $\rho_{0}$ (homogeneous) \\ \hline
  \end{tabular}
\label{EBHM_GHM}
\end{table}

By using the GP equations for the EBHM, we investigated dynamical properties 
of the electric flux put on a line as an initial state.
In the confinement region of the GHM, the string breaking and restoration of 
the electric flux is observed, which has similar properties to the Schwinger
mechanism in QED.
On the other hand in the Higgs region, the flux spreads out and many bits of 
electric flux develop.
Finally we discussed the reliability of the results obtained by the GP equations
by using the idea of the TWA, and gave some remarks for the experimental 
setup to observe the string-antistring oscillation.
We hope that the above phenomena will be observed experimentally in the near future.
In particular, dynamical behavior of the transition from the confinement to Higgs phases 
is quite interesting and expected to be observed by the strong controllability 
and the versatility of the cold atomic system.

Finally, let us note that
our approach identifies the phase and the amplitude of atomic variable
as U(1) vector potential and the electric field respectively,
and introduce the Higgs field to relax the crucial fine tuning to
realize the Gauss' law.
These general and universal procedures make us possible to  
define corresponding GHMs of wide variety
both for high and low fillings and also for homogeneous and 
inhomogeneous atomic distributions.
Although this  GHM may contain nonlocal interactions in some cases
and look quite different from the conventional lattice gauge theory,
it is important that GHM is of course a ``gauge theory having gauge symmetry".
It obviously opens the way to study gauge theory in more general view points.
Also such gauge-theoretical view points may shed some lights backward
on studying the original atomic systems as partly shown in Sec.~\ref{interp}.

\acknowledgments
Y. K. acknowledges the support of a Grant-in-Aid for JSPS
Fellows (No.JP15J07370). This work was partially supported by Grant-in-Aid
for Scientific Research from Japan Society for the 
Promotion of Science under Grant No.JP26400246, JP26400371 and JP26400412.

\appendix
\renewcommand{\thefigure}{\Alph{section}.\arabic{figure}}
\setcounter{figure}{0}
\renewcommand{\theequation}{A.\arabic{equation}}

\section{Global phase diagram of extended Bose-Hubbard model 
in the $(V/J-U/J)$ plane for large fillings}\label{appa}

In this Appendix, we study the global phase diagram of the 1D EBHM 
in the \textit{whole} ($V/J$-$U/J$)-plane at large fillings 
by taking account of both 
{\it homogeneous and inhomogeneous local density average} $\rho_{0a}
\equiv \la \hat{\rho}_a \ra$.
To this end, we first derive an effective model, which is to be used for
study of the EBHM in the whole parameter region in the ($V/J$-$U/J$)-plane,
in particular even for a negative $g^2$. 
This model includes the 
local variational density introduced in Ref.~\cite{Kuno}
in addition to the spatial vector potential $\theta_{x,1}$,
and is capable to cover the case of inhomogeneous 
local density average for $g^2 \ll 0$ right region of Fig.~\ref{Figdw1d}.
This model is complement to the gauge-theoretical study of the EBHM 
for $g^2>0$ in Sec.~\ref{relation} as we explain later on.
In this Appendix, we also show the results of MC simulation of this effective model, and
compare the result with that of Sec.~\ref{phasediagram}.

\subsection{Effective model for homogeneous and inhomogeneous local density 
with large fillings}

In contrast to Eq.(2), we consider a general 
($a$-dependent) local density average $\rho_{0a}$
small fluctuations around it by writing
\be
\hat{\rho}_a = \rho_{0a} +\hat{\eta}_a.
\ee
To derive the effective model, we expand $H_{\rm EBH}$
with respect to $\hat{\eta}_a$ up to $O(\hat{\eta}^2)$
and discard the term like $\Delta H$ in Eq.~(\ref{HGH})  
to obtain
\be
H_{\rm EBH} &=& H_{\rho_{0a}} +H'+ O(\eta^3),\nn
H_{\rho_{0a}}&=&\sum_a\left[{U\over 2}
\rho^{2}_{0a}+V\rho_{0a}\rho_{0,a+1}
-\mu\rho_{0a}\right],\nn
H'&=& \sum_a\left[\frac{V}{2}(\eta_{a+1}+\eta_a)^{2}+
\frac{g^2}{2}\eta^{2}_a\right.\nn
&&\left.-2J\sqrt{\rho_{0,a+1}\rho_{0a}}\cos(\theta_{a+1}-\theta_a)\right], 
\label{HGH2}
\ee
where the suffices $a$ of the variables refer to sites in the optical lattice.
We have included
the chemical-potential term in the last term of $H_{\rho_{0a}}$
to fix the average total particle number
 $\la \sum_a\hat{\rho}_a\ra=\sum_a \rho_{0a}=L\rho_0$.
Because we treat $\{\rho_{0a}\}$ as the variational parameters, 
the partition function $Z_{\rm EBH}={\rm Tr} \exp(-\beta H_{\rm EBH})$ is expressed as 
\be
Z_{\rm EBH}&=&{\rm Max}\ Z_\rho(\{\rho_{0a}\}),\nn
Z_\rho(\{\rho_{0a}\})&=&\int[d\eta][d\theta]\exp\left[
 \sum_{a,\tau}\left(-i\eta_a\nabla_0\theta_a -\Delta \tau
 H_{\rm EBH}\right)\right],\nn
 \label{ZEBH}
 \ee
where the symbol `Max' in the first line indicates to find the maximum value 
with respect to the parameters $\{\rho_{0a}\}$. (For details, see Ref.~\cite{Kuno}.)
As in Sec.~\ref{relation}, all the fields {\it except $\rho_{0a}$} are defined
on the (1+1)D lattice with the site $(\tau,a)$ as $\eta_a(\tau), \theta_a(\tau)$
(the argument on $\tau$ in these fields is omitted)
and $\nabla_0\theta_a\equiv \theta_a(\tau+1)-\theta_a(\tau).$

First, we consider the case of $g^2 < 0$.
As  we see in Sec.~\ref{relation}, the density fluctuations $\eta_a$ 
in $H'$ of Eq.~(\ref{HGH2}) become unstable for $g^2 < 0$.
This {\em instability} is related to the appearance of phase such as the DW.
However as the mean-field theory in Sec.~\ref{relation} shows, 
the condition $g^2 < 0$ does not necessarily 
give the DW phase due to the effect of the hopping term.
Therefore, more careful study is required for the case $g^2<0$.

Returning to the original EBHM, {\em higher-order terms of $\eta_a$ appear 
in the potential from the hopping term, which prefers the homogeneity of the system}.
This is the origin of 
the `homogeneous phase' between the region $g^2>0$ and the inhomogeneous
region shown in Fig.~\ref{Figdw1d}.
Then, let us see what happens if the
higher-order terms of $\eta_a$ up to $O(\eta^4)$ are taken into account.
The potential energy 
$V(\eta)$ for a configuration $\eta_a=(-)^a\eta$ has the following form,
\be
V(\eta)&=& \alpha g^2 \eta^2 +\lambda \eta^4 \ (\alpha ,\lambda > 0)\nn
&=&\lambda(\eta^2-\xi^2)^2, \nn
\xi&\equiv& \sqrt{\frac{-\alpha g^2}{2\lambda}}\ (>0\ {\rm for}\ g^2<0),
\label{veta}
\ee
where $\eta^3$-term is deleted by the optimal choice of $\{\rho_{0a}\}$ and 
we put $\theta_{a+1}=\theta_a$ to minimize the hopping term in the 
Hamiltonian as we are mostly interested in the phase boundary of the SF.
The potential $V(\eta)$ has a double-well shape for $g^2<0$ and the potential minima
are given at $\eta_a=\pm \xi$; 
where $\xi (>0)$ is a certain number that depends on 
the parameters $J, U, V$ and $\rho_0$.
The above observation suggests an approximation that one takes only 
those two configurations $\eta_a =\pm \xi$
into account instead of integrating over $\eta_a \in (-\infty,\infty)$ \cite{etasum}.
If the amplitude of the fluctuation $\xi$ is (sufficiently) small, 
the homogeneous configuration is realized, which occurs 
for moderate values of U, V, and negative $g^2=(U-2V)$
\cite{etaorder}.
Here, it should be remarked that {\em this `phase' is reminiscent of 
the Haldane insulator} \cite{DMRG1,Kawaki}, which is observed for the EBHM at low fillings,
although the non-local string order does not exist in the present case.

To perform the $\eta$-integration in the path integral,
we make further approximation by factorizing integrations over $\{\eta_a\}$'s
to the product of single-site integrals over each $\eta_a$.
This implies that we neglect correlations among $\{\eta_a\}$ described by the  
NN interactions of the $V$ and $J$ terms in $H'$ of Eq.~(\ref{HGH2}).
By using these approximations, we evaluate $Z_\rho$
in Eq.~(\ref{ZEBH}) for the case of $g^2 < 0$ as follows;
\be
Z_\rho\!&=&\!\!\int [d\theta_1]Z_\theta \exp\!\left[-2J\Delta \tau
\!\sum_{a,\tau}\!\sqrt{\rho_{0,a+1}\rho_{0a}}\cos(\theta_{a+1}\!-\!\theta_a)\right],\nn
Z_\theta\!&=\!&\int\! [d\eta]\exp\!\left[\Delta\tau\sum_{a,\tau}\Big( -i\eta_a
\nabla_0\theta_a- \frac{g^2}{2}\eta_a^2+D_4(\eta^4)\Big)\right] \nonumber \\
&
\simeq& \prod_{a,\tau}\left[\exp(i\Delta\tau\xi\nabla_0\theta_a)
+\exp(-i\Delta\tau\xi\nabla_0\theta_a)\right] \nonumber \\
&\simeq& \exp\left[-\frac{1}{2}\sum_{a,\tau}(\xi\Delta\tau)^2 (\nabla_0\theta_a)^2\right]\nonumber \\
&\simeq& \exp\left[\sum_{a,\tau}(\xi\Delta\tau)^2\cos(\nabla_0\theta_a)\right],
\label{sumeta}
\ee
where $D_4(\eta^4)$ refers to the term of $O(\eta^4)$, which was used to obtain 
Eq.~(\ref{veta}).
The approximate expression for $Z_\theta$ is the result discussed above, and
we have incorporated the compactness of $\theta_a$ in the final expression.

From Eq.~(\ref{sumeta}), an effective theory of the EBHM
for the case $g^2<0$, which we call $Z^{\rm E}_\rho$, 
is derived as in the previous discussion in Sec.~\ref{relation}, 
\begin{eqnarray}
&&Z^{\rm E}_{\rm EBH}={\rm Max}\ Z^{\rm E}_\rho(\{\rho_{0a}\}),\nn
&&Z^{\rm E}_\rho(\{\rho_{0a}\}) = \int\prod [d\theta]\exp(A_{\rm eff}
-\Delta\tau H_{\rho_{0a}}), \label{effective1} \\
&&A_{\rm eff}=\sum_{a,\tau}\Big[C_\tau
\cos\nabla_0\theta_a+2J\Delta\tau\sqrt{\rho_{0,a+1}\rho_{0a}}
\cos(\nabla_1\theta_a)\Big], \nonumber
\end{eqnarray}
where the parameter $C_\tau$ is given as $C_\tau=2(\xi\Delta\tau)^2$ 
from Eq.~(\ref{sumeta}) {\em for the case $g^2<0$}.
The above model defined by Eq.~(\ref{effective1}) is the effective theory,
which is used for study on the global phase diagram of the EBHM as we discuss
below. 
\begin{figure*}[t]
\centering
\hspace{-1cm}
\includegraphics[width=15cm]{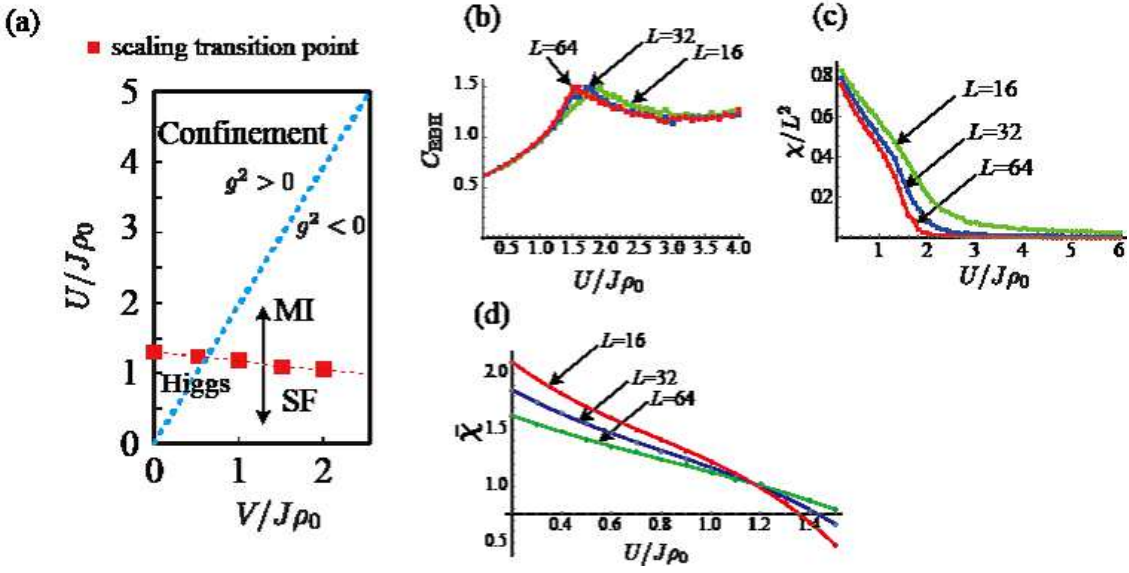}
\caption{(Color online) (a) Phase diagram of  the effective model
in Eq.(\ref{effective1}) obtained by the MC simulations. 
This MC calculation {\em includes the update of the variational parameter} 
$\{\rho_{0a}\}$. 
The phase boundaries determined by the scaling
of $\chi=\sum_\ell G_{\rm B}(\ell)$. 
(b) Specific heat $C_{\rm EBH}$ for $V/J=1$. 
(c) Susceptibility density $\chi/L^{2}$ in three different lattice size $L=16,32$ and $64$.
(d) Scaled susceptibility density $\bar{\chi}=L^{\eta-2}\chi$.
To determine the detailed phase boundary of the SF phase,  we use
$\bar{\chi}$. 
By choosing the exponent $\eta$ and $U/J$ suitably, we found that three curves of $\bar{\chi}$ 
merge at a point, which is the transition point (see Sec. III).
Through this analysis, we verified that the SF-MI transition is a BKT-type phase transition. 
For $V/J=\rho_0$, the critical point is $U/J\sim 1.17\rho_{0}$
and the exponent is $\eta\sim0.24$.
MC simulation of the effective model in Eq.(\ref{effective1}) under-estimates
the SF in the vicinity $g^2\approx 0$ as explained in the text.
}
\label{fig.PD3}
\end{figure*}

Let us consider {\em the case $g^2 > 0$}.
As we shall see,  the effective model $Z^{\rm E}_\rho$ in Eq.~(\ref{effective1})
can be also used to calculate the phase diagram for $g^2 > 0$ 
if  the constant $C_\tau$ is suitably chosen.
This is because 
$A_{\rm eff}$ in Eq.~(\ref{effective1}) has the same structure with
$A_{\rm GH}$ of Eq.~(\ref{AGH}) in the unitary gauge $\phi_x=1$ for {\em small $V$}. 
In fact,  $A_{\rm I}$ in Eq.~(\ref{AGH}) squeezes $\theta_{x,0}\to 0$ (mod $2\pi$)  
for $V \to 0$.
Then, $A_{\rm P}$ reduces to the $C_\tau$-term in Eq.~(\ref{effective1})
[To make comparison, we recall the relation $\theta_{x=(\tau,r),1}=(-)^r\theta_{a,1}(\tau)$].
$A_{\rm H}$ reduces to the second term of Eq.~(\ref{effective1}) with homogeneous $\rho_{0a}$
which is preferred by $H_{\rho_0}$ for small $V$.
Namely, two system coincide with the  following relations; 
\begin{eqnarray}
&& A_{\rm P}\rightarrow  c_2 \cos(\nabla_0\theta_{a,1})
\leftrightarrow C_\tau \cos(\nabla_0\theta_{a,1}),  \nn
&& A_{\rm H}\rightarrow c_3 \cos(\nabla_1\theta_{a,1})
\leftrightarrow 2J\Delta\tau\rho_0 \cos(\nabla_1\theta_{a,1}),\nn
&&C_\tau=c_2\equiv \frac{1}{g^2\Delta\tau}\simeq \frac{1}{U\Delta\tau}, \ 
c_3= 2J\Delta\tau\rho_0,
\end{eqnarray}
where the last relation is the same as Eq.~(\ref{cs}).

Finally for sufficiently large $V (>U\gg J)$, the effective theory in 
Eq.~(\ref{effective1}) correctly predicts the existence of the DW.
In fact 
for this parameter region, $H_{\rho_{0a}}$ in Eq.~(\ref{effective1}) is dominant, 
which favors the DW state, i.e., the state with 
$\rho_{0a}=\rho_0+(-)^a\delta\rho/2 $, as the ground-state. 
The other terms related to the quantum fluctuations $\theta_a$ are
irrelevant to the formation of the DW whose existence is governed by 
the variational slow variables $\{\rho_{0a}\}$.

Here, we should remark on the applicability of the effective model Eq.~(\ref{effective1}).
In the discussion for the case $g^2<0$, we simply put $\eta_a=\pm \xi$
on the integration over $\eta_a$.
Also for the case $g^2>0$, only in the region $U\gg V$, $A_{\rm eff}$
coincides with $A_{\rm GH}$, in which the effects of the NN repulsion
(i.e., the $V$-term) are taken into account properly.
Therefore, $A_{\rm eff}$ in Eq.~(\ref{effective1}) is not applicable
for the region $g^2\approx 0$ although the MC study of it gives
suggestive results in that region as, $\{ \rho_{0a}\}$ are treated 
as variational parameters in $A_{\rm eff}$.
See the results in the following subsection.

In conclusion, the considerations in this subsection support to use the model
defined by Eq.~(\ref{effective1}) to study the EBMH.

\subsection{Global phase diagram in the $(V/J-U/J)$-plane for large fillings}

In this subsection we calculate the global phase diagram in the $(V/J-U/J)$-plane
of the 1D EBHM  for large fillings by performing MC simulation of the effective theory of Eq.(\ref{effective1}).

In the practical calculation, we put $\Delta\tau=1$, 
and the global average $\rho_0$ as the unit. 
We also put $C_\tau=1/U$ as the on-site repulsion hinders the SF.
This manipulation may underestimate the region of the SF in the phase diagram 
since in certain parameter region of $g^2\approx 0$, fairly large density fluctuations 
exist and as a result, the enhancement of the SF may take place \cite{kawaki}.
Phase boundaries are determined by calculating 
``internal energy" $U_{\rm EBH}$ and ``specific heat" $C_{\rm EBH}$ defined by 
\begin{eqnarray}
U_{\rm EBH}&=&-\frac{1}{L^2}\langle A_{\cal T}\rangle,\\
C_{\rm EBH}&=&\frac{1}{L^2}\langle (A_{\cal T}-\langle A_{\cal T}\rangle)^2\rangle.
\\
A_{\cal T}&=&A_{\rm eff}-\Delta\tau H_{\rho_{0a}}, \nonumber
\end{eqnarray}
where the above quantities are calculated using the effective model 
Eq.(\ref{effective1}).
We also use  the boson correlation function $G_\psi(\ell)$
and the DW correlation function $G_\rho(\ell)$ defined by
\begin{eqnarray}
G_{\psi}&=&\langle\cos( \theta_{a+\ell}-\theta_{a})\rangle,\nn
G_{\rho}&=&(-1)^\ell\langle(\rho_{0,a+\ell}-\rho_0)(\rho_{0a}-\rho_0)\rangle,
\label{Grho} 
\end{eqnarray}
to identify the phases.
\begin{figure}[t]
\centering
\includegraphics[width=7.5cm]{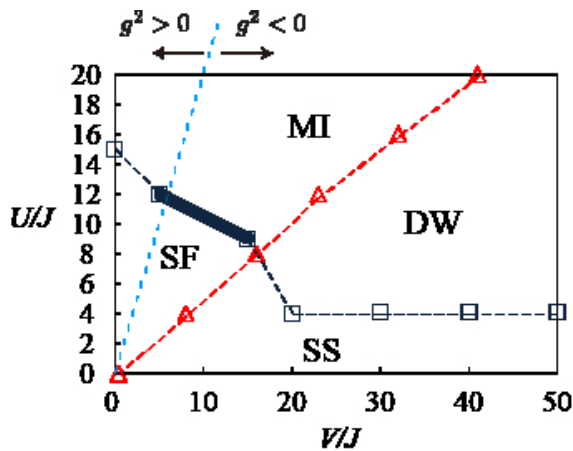}
\caption{(Color online) 
Phase diagram of  the effective model
in Eq.(\ref{effective1}) obtained by the MC simulations for $L$=16. 
Average filling is 10 per site ($\rho_{0}=10$) and we employed canonical ensemble.
This MC calculation {\em includes the update of the variational parameter} 
$\{\rho_{0a}\}$. The target regime is large $V$.
The specific regime of the phase boundary between SF and MI, $5\lesssim V/J\lesssim10$ is not clear.}
\label{fig_largeV}
\end{figure}

We employed the standard Metropolis algorithm \cite{Met} with the local update as before.
{\em To begin with, we fix the variational parameters}, i.e., employ the assumption of
homogeneous density $\rho_{0a}=\rho_{0}$. This manipulation corresponds 
to the London limit  for a U(1) Higgs field $\phi_r=\exp(i\varphi_r)$ in Sec.~\ref{relation}. 
In small $V$ ($0<V/J\lesssim 2$), this assumption is justified because the DW order is 
not expected there.
This manipulation allows us to compare the obtained phase diagram with that
of the GHM. The result is shown in Fig.\ref{fig.PD2}. 

Next, the phase diagram {\it including a slow update of the variational 
parameter $\rho_{0a}$} in small $V$ ($0<V/J\lesssim 2$), is shown in Fig.\ref{fig.PD3}. 
Some quantities used for location of the phase boundaries are shown in 
Fig.\ref{fig.PD3} (b), (c) and (d).
This phase diagram also has a similar structure to that of the GHM for $g^2>0$.
However, the phase diagram obtained by the MC of the GHM exhibits
an enhancement of the SF, i.e., the Higgs phase, in the vicinity of $g^2=0$.
As we explained above, this result comes from the fact that strong fluctuations
of $\{\eta_a\}$ in that parameter region are taken into account properly in 
the GHM whereas not in the derivation of $A_{\rm eff}$.
Furthermore even for $V\to$ small, the phase transition lines 
in the two models do not coincide.
Reason for this discrepancy is that in the $Z^{\rm E}_{\rm EBH}$ 
in Eq.(\ref{effective1}), the local density fluctuations $\{\rho_{0a}\}$ are
taken into account.
In fact, this hinders the effective hopping amplitude 
$J\sqrt{\rho_{0,a+1}\rho_{0a}}$ as the function $f(x)=\sqrt{x(1-x)}$ has the 
maximum at $x=1/2$.

Finally, Fig. \ref{fig_largeV} exhibits a phase diagram 
{\it without rescaling by the mean density $\rho_0$}.
Here, we put, as an example of the large filling, $\rho_{0}=10$ 
(i.e., average occupation numbers of atoms per site is ten)\cite{filling}.
The coupling constant $J$, $U$, and $V$ are not rescaled 
by the mean density.
We found that there are four phases, SF (superuid), MI (Mott insulator), 
DW (density wave), and SS (supersolid). 
The DW state is recognized by the AF-type staggererd
configuration of the local density average $\{\rho_{0a}\}$.
It is possible that both the quasi-long-range orders of the SF and the diagonal DW 
order coexist, indicating the supersolid (SS) state.
The phase boundaries between SF(SS) and MI(DW) 
are determined by $\chi=\sum_\ell G_{\rm B}(\ell)$. 
On the other hand, the boundaries between SF(MI) and MI(SS) is 
determined by the specific heat peak.

\subsection{Gauge theoretical interpretation including inhomogeneity of density}\label{interp}

Let us consider the physical meaning of each phase of the EBHM in the terminology 
of gauge theory \cite{Kogut}.
The SF phase corresponds to the Higgs phase as the gauge fields 
$\theta_{r,\mu}$ has small fluctuations there. 
On the contrary, the MI corresponds to the confinement phase,
as the boson density $E_r$, which is the electric field in the GHM, has
small fluctuations in that phase; 
therefore, the vector potential $\theta_{r,\mu}$ fluctuates strongly in the MI.
The above identification was verified in Sec.~\ref{dyn} through the study 
on the real-time dynamics of the electric flux.

Next, we consider the DW and SS phases which appear for $g^2 \ll 0$. The DW phase has a definite density imbalance between even and odd sites., so that the density fluctuation from the mean density is basically small. Therefore, in terms of the gauge theory, the fluctuation of the electric field is small, which means that the DW phase follows the property of the confinement phase. We can say that the DW phase does not have substantial difference from the MI phase from gauge-theoretical point of view.  
In this sense, the SS state corresponds to the ordinary Higgs phase as the gauge fields $\theta_{r,\mu}$ have small fluctuations.
In Fig.~\ref{fig.PD3} and Table \ref{EBHM_GHM2}, we show 
the gauge-theoretical interpretation of the phase diagram of the EBHM.

\begin{table}[h]
\caption{ Correspondence between the phases of EBHM and GHM. 
$\Delta\theta$ is the magnitude of fluctuations of atomic phase $\theta_a$ 
(vector potential $\theta_{r,1}$).}
  \begin{tabular}{|c|c|c|c|} \hline
    EBHM & U(1) GHM & $\Delta \theta$ & $\rho_{0a}$  \\ \hline
    MI &  Confinement & large & $\rho_{0}$ (homogeneous)\\ 
    DW & Confinement & large & $\rho_{0,{\rm even}}\neq \rho_{0,{\rm odd}}$ \\ 
    SS & Higgs & small &  $\rho_{0,{\rm even}}\neq \rho_{0,{\rm odd}}$ \\ 
    SF & Higgs & small &  $\rho_{0}$ (homogeneous) \\ \hline
  \end{tabular}
\label{EBHM_GHM2}
\end{table}


\begin{thebibliography}{99}
\bibitem{advantage}
Some further advantages of the present approach of identifying the phase
 of atomic variable to the U(1) gauge field over the quantum link-model is discussed in depth 
 in Sec.~I of Ref.~\cite{Future}.
\bibitem{etaorder}
In the present discussion, we assume that a long-range order 
of the Ising-type fluctuations of $\eta_a$ does not exist.
If this long-range order exists, the system is in the density-wave. 
In this phase, the finite `magnetization' exists.
\bibitem{kawaki} This enhancement of the SF is actually observed by the
SSE-quantum MC simulation of the EBHM at low fillings.
See Ref.~\cite{Kawaki}
\bibitem{filling}
In general, it is difficult to carry out a numerical simulation for large filling  beyond mean-field theory due to enormous consumption of computer memory.
\end{thebibliography}
\end{document}